\documentclass[aps,showpacs,showkeys,twocolumn,floatfix]{revtex4}

\usepackage{graphicx}

\begin{document}

\title[VLTI AT polarimetry]{Calculated Ellipsometry of the VLTI AT Mirror Train}

\author{Richard J. Mathar}
\homepage{http://www.strw.leidenuniv.nl/~mathar}
\email{mathar@strw.leidenuniv.nl}
\affiliation{Leiden Observatory, P.O. Box 9513, 2300 RA Leiden, The Netherlands}


\date{\today}

\begin{abstract}
The polarization effect of the 31 reflections within the
mirror train
of an auxiliary telescope of the Very Large Telescope Interferometer
is calculated as a function of pointing direction and rotator angle.
With a rough estimate of the mean complex index of 
refraction of the reflecting surfaces, their Jones matrices are concatenated
while tracing a ray from M1 up to the feeding mirror in front of an instrument.
The net effect is summarized in terms of the axis ratio of the polarization
ellipse of star light that was circularly polarized above the primary mirror.
\end{abstract}

\maketitle
\section{Coude Train of Auxiliary Telescopes}

This manuscript is concerned with the polarimetric characterization of the mirrors relevant
to observations with the Very Large Telescope Interferometer (VLTI). The competition
for observing time with the 8-m Unit Telescopes implies that the 1.8-m Auxiliary Telescopes (AT's)
are more important to interferometry from a statistical point of view, and only these will
be considered here.

\begin{figure}[hbt]
\includegraphics[scale=0.6]{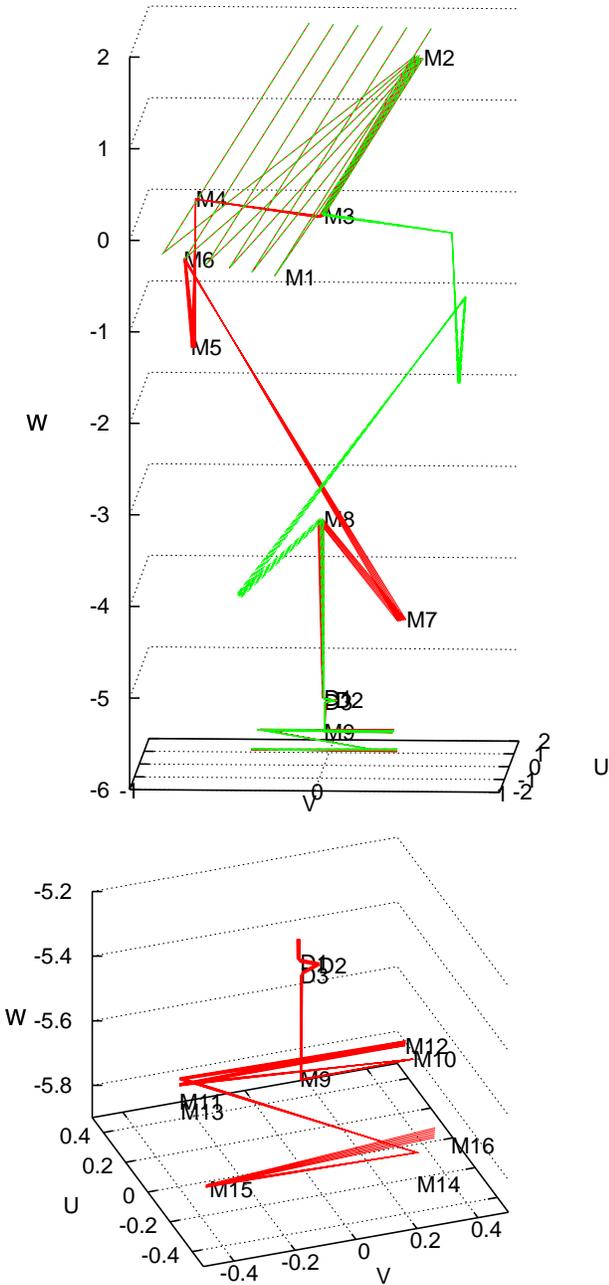}
\caption{
The AT mirror cascade.
Configurations with two different locations of the Nasmyth platform,
which place mirrors M4 to M7 at opposite sides of the vertical azimuth axis,
are shown in red (N+) and in green (N-).
The magnified view
below is painted with one color only;
in this section the geometric paths of N+ and N- do not differ.
}
\label{fig.ATfig1}
\end{figure}

The results demonstrate the polarization inherent to one beam delivered
by one telescope.
The more challenging application to compute
the \emph{differential} polarimetry differentiating two beams by two different telescopes \cite{BuscherSPIE7013}
is not addressed; in the sense that coherency remains
intact supposing the phases in the two polarization sub-channels are modified
by the same amount in both beams, interferometry is not threatened.
However, if an interferometric instrument depends on a scheme
that measures the fluxes in the two polarizations separately,
imbalanced fluxes
induce unfavorable signal-to-noise ratios
\cite{DelplanckeNAR52,SahlmannAA507}.

The configuration of the reflecting mirror optics is outlined in Figure \ref{fig.ATfig1} \cite{DelrezLiege36},
which traces five rays at various distances to the optical axis starting with the first reflection off
the 1.8-m primary M1:
\begin{itemize}
\item
8 reflections in the Coud\'e train.
(The design of the unit telescopes \cite{BeddingIAU158} up to M8 is similar.
The first three of these appear to be understood \cite{WitzelAA525}.)
Incidence angles at M3 and M4 are typically close to 45$^\circ$;
\item
3 reflections (D1, D2, D3)
inside a star rotator \cite{DixonAO18}
with incidence angles of 19$^\circ$ and two times $54.5^\circ$;
\item
and 8 reflections M9 to M16 inside the star separator \cite{DelplanckeSPIE5491},
incidence at M9 close to $45^\circ$ and otherwise $<16^\circ$.
\end{itemize}
After these first 19 reflections, the light beam leaves M16 and heads with
a nominal diameter of 80 mm into the
light duct.
The followup reflections are included taking
rough information on the geometries (conic constants and apex positions)
with a ruler from a variety of design sketches
\cite{WallanderSPIE4848,FerrariSPIE4006,GittonSPIE4838}:
\begin{itemize}
\item 1 reflection at $45^\circ$ off what was M12 in the nomenclature before introduction of
the star separator \cite{GuisardSPIE4838} to align the beam with the $\pm U$ axis of the delay line tunnel;
\item 5 reflections inside the main delay line cat's eye \cite{HogenhuisSPIE4006};
incidence typically $<5^\circ$;
\item 1 reflection at $45^\circ$ pushing the beam from the delay line tunnel into the VLTI laboratory, into the $+V$ direction,
with the aid of M16 in the old nomenclature \cite{SchoellerNAR51};
\item 3 reflections below $4^\circ$ inside the beam compressor, quenching the beam diameter to 18 mm;
\item 1 reflection in the ``switch yard'' with incidence angle $45^\circ$
either forward to the differential delay lines or to the instruments
(The 5 reflections inside differential delay lines are only relevant to a
mode of one instrument and not included for that reason \cite{LaunhardtASP338,PepeSPIE7013});
\item 1 dichroic reflection at $45^\circ$ into the $+V$ direction;
feeding individual instruments.
\end{itemize}

Numbers in Section \ref{sec.res} are a snapshot including these 31 reflections,
defining the interface towards the first optical element of a generic VLTI instrument.

\begin{figure}[hbt]
\includegraphics[scale=0.35]{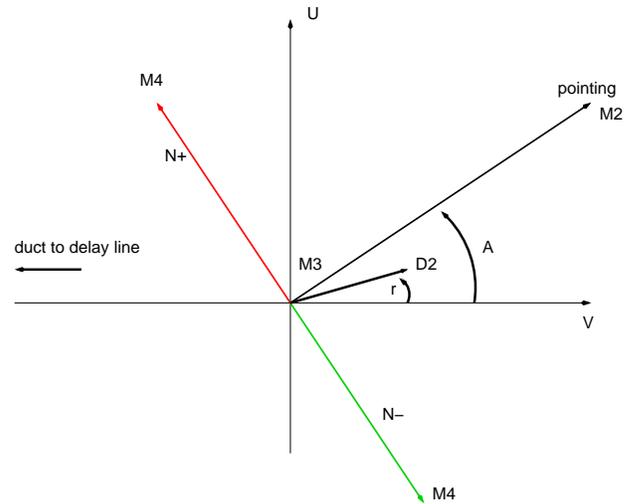}
\caption{
The coordinates in a bird's view, looking down the
negative $W$ direction of Figure \ref{fig.ATfig1}. M8, D1, D3 and M9 appear stacked below M3.
The azimuth $A$ is defined relative to the direction in which the beam
leaves the Coud\'e train into the duct. The two prospective directions
from M3 to the  Nasmyth focus and to M4 co-rotate with the vertical axis, $A+90^\circ$
for N+, $A-90^\circ$ for N-.
}
\label{fig.coord}
\end{figure}

Figure \ref{fig.coord} clarifies the azimuth definitions of this manuscript
in a projection looking from the zenith above the telescope down the vertical
axis onto M3 or M8\@. The position of the
star fixes the angle $A$ and zenith angle $z$ in local horizontal
$V$, $U$, $W$ coordinates. The rotator angle $r$ (which rotates
the image by $2r$) is defined with the same azimuth reference, characterized
by the position of D2\@.

\section{Ray Tracing} 
\subsection{Reflective Surface Ellipsometry}\label{sec.n}

Given the incidence angles on the mirror surfaces, the polarimetric
effects on the beam can be calculated ``in principle'' (to quote Tinbergen \cite{TinbergenPASP119})
if the complex valued refractive index $n$ of these are available
at the wavelength of interest.

\begin{figure}[htb]
\includegraphics[scale=0.65]{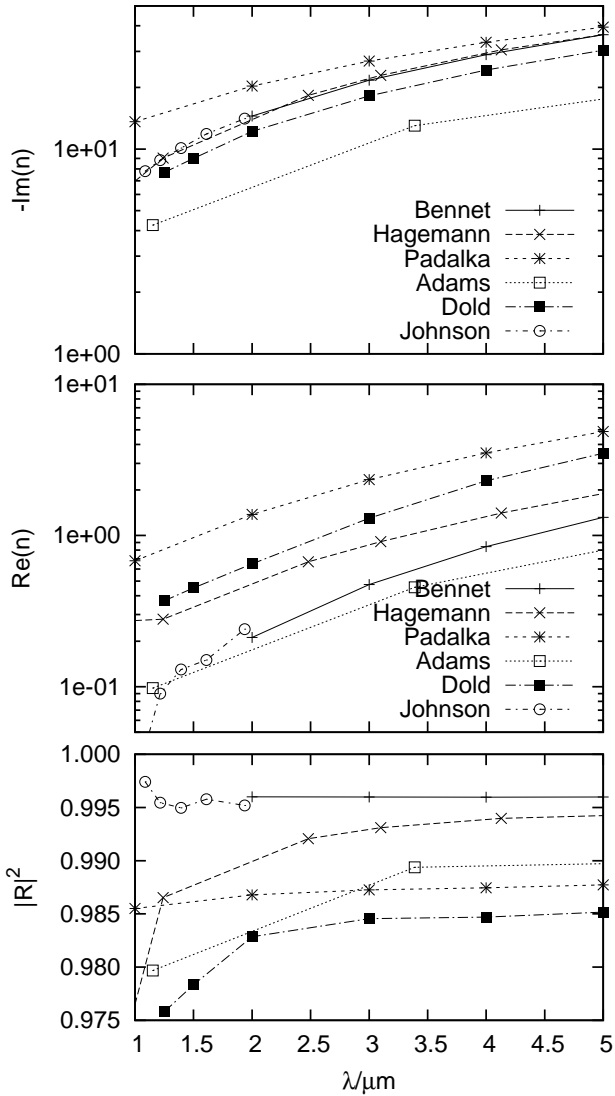}
\caption{
Middle and top: Real part and absolute value of the imaginary part of measured
refractive indices $n$ of Silver \cite{DoldO22,PadalkaOS11,
Bennet1966,AdamsOptComm15,HagemannJOSA65,JohnsonPRB6} as a function
of wavelength $\lambda$.
Bottom: associated intensity reflection coefficient at zero incidence,
$|R|^2=|(n-1)/(n+1)|^2$.
\label{fig.hagemann}}
\end{figure}

The complex-valued index of refraction for Ag films in the infrared is discussed
in the literature \cite{DoldO22,PadalkaOS11,SchulzJOSA44_357,
SchulzJOSA44_362,SchulzJOSA44_540,BennetPR165,AdamsOptComm15,HagemannJOSA65,OrdalAO22},
summarized
in Fig.\ \ref{fig.hagemann}\@.
Selecting a refractive index out of these is an obvious source of large error
in all results shown further down, and the largest
uncertainty in modeling efforts of this kind \cite{BeckAAp443}.
Early design assumed values $|R|^2 > 0.98$ \cite{MerkleESO24,BurgeAO12}.

Because no ellipsometric information
on any of our reflections is available, we start from a constant
$n=0.67-18.3i$ for all surfaces, which is the Hagemann--Gudat--Kunz
value of Silver interpolated to a wavelength of 2.48 $\mu$m on
logarithmic scales of the real and imaginary part \cite{HagemannJOSA65}.
We ignore that M1 is clearly aluminized \cite{EttlingerMess97} and that M9 is a dichroic.
The net effect of this choice is an energy transmission coefficient
(absolute value of the determinant of the Jones matrix or the $M_{II}$ element
of the equivalent Mueller matrix) near 0.77\@.

In a desperate act of tuning
this to the anticipated transmission coefficient near
0.2---geometrically scaled
by mirror count to a different configuration \cite{vlt150001826} and removing
an estimated contribution from  diffuse reflection---,
this value of $n$ is multiplied by 0.13. The calculations actually use
$n=0.0871-2.379i$. The average intensity reflection coefficient at the
average incidence angle is brought down to $\sqrt[31]{0.2}\approx 0.95$.

\subsection{Jones Calculus} 
The ray tracing is started with a minimum distance to the optical axis, as enforced
by the M2 diameter. A small corrugation of polarization across the beam as a function
of the incidence point on M1 is induced by the variation of incidence angles
on the mirror curvatures
\cite{BreckinridgeApJ600,PatatPACS118,ShamirJModOpt38}.
It is not studied here.

The Fresnel coefficients or each reflection are the two entries on
the diagonal of each Jones matrix \cite{HurwitzJOSA31,JonesJOSA31_488,JonesJOSA32,PistoniAO34,EliasAJ549},
in the $s-$ $p-$basis for the
two states of polarization.
The contribution of an individual reflection is the product with the $2\times 2$ rotation matrix
that adapts the plane of incidence of the previous mirror surface to the plane
of incidence at the surface point of the next mirror, and with the free flight propagation term
which is a diagonal matrix with two phasors depending on the (optical)
path length of the ray from one mirror surface to the next \cite{McGuireAO33,FymatAO10_2711}.

\begin{figure}[hbt]
\includegraphics[scale=0.35]{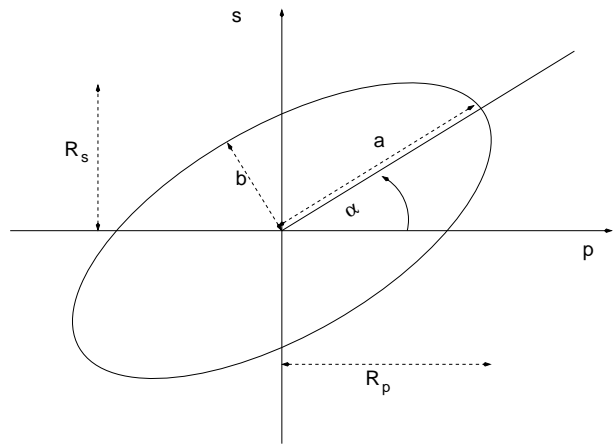}
\caption{Illustration of the main ellipsometric variables that
characterize the polarization state.
}
\label{fig.elli}
\end{figure}

The net effect of the mirror train train is the product
of the matrices, four complex values $J$.
It converts an initial Jones vector with
amplitudes $A_{p,s}$ into a vector with amplitudes $R_{p,s}$
(Figure \ref{fig.elli}),
\begin{equation}
\left(
\begin{array}{cc}J_{pp} & J_{ps} \\ J_{sp} & J_{ss}
\end{array}
\right)
\cdot \left( \begin{array}{c} A_p \\ A_s\end{array} \right)
=
\left( \begin{array}{c} R_p e^{i\delta_p}\\
R_s e^{i\delta_s}\\
\end{array} \right)
.
\end{equation}

The values plotted in figures \ref{fig.pol_0}--\ref{fig.pol_60} consider circularly polarized
light with
\begin{equation}
\left(\begin{array}{c}A_p\\ A_s\end{array}\right)
=
\left(\begin{array}{c}1\\i\end{array}\right)
\label{eq.sigma}
\end{equation}
as the input above M1.
The relative phase shift is $\delta \equiv \delta_s-\delta_p$ \cite{Born2}.
The ellipse main axis is rotated
by an angle $\alpha$ relative
to the main coordinates (where $A_p$ is the horizontal component
as we consider the beam in the interferometric laboratory where the 
planes of incidence are horizontal):
\begin{equation}
\tan\alpha = \frac{R_s}{R_p}.
\label{eq.alph}
\end{equation}
This formula defines $0\le \alpha \le 90^\circ$ as the direction of the longer
or of the shorter axis. One can attach it uniquely
to the longer axis if the sign of $R_p$ is flipped and
$\delta_p$ shifted by 180$^\circ$ for all cases where $\cos\delta < 0$.
This will add 90$^\circ$ steps
in the followup graphs of $\alpha$ wherever the ellipse passes
through
a ``degenerate'' circle as a function of pointing direction.

Let  the amplitudes of the ellipse in its main coordinates be $a$ and $b$
(Fig.\ \ref{fig.elli}).
The amplitude (axis) ratio of the polarization ellipse then is
\begin{equation}
\tan\chi = \frac{b}{a},
\end{equation}
which is computed from $\alpha$ and $\delta$ via \cite{Born2}
\begin{equation}
\sin(2\chi)=\pm \sin(2\alpha)\sin\delta.
\end{equation}

The diattenuation
\begin{equation}
P\equiv \left|\frac{b^2-a^2}{a^2+b^2}\right|
=
|\cos(2\chi)|
\label{eq.P}
\end{equation}
measures the asymmetry of the ellipse, converted from field
amplitudes to
energy or flux.
(In Figure \ref{fig.elli}, $b/a\approx 0.44$ and therefore $P\approx 0.67$.)
The value of $P$ turns out to be
the same for the left- and right-handed circulation of the circular polarization,
that is, to be immune against flipping $i$ to $-i$ in the lower
component in (\ref{eq.sigma}).
The value of $\alpha$ changes by 90$^\circ$ under this reversal of the input polarization.

\section{Results}\label{sec.res}
\subsection{Diattenuation}

\begin{figure}[hbt]
\includegraphics[scale=0.65]{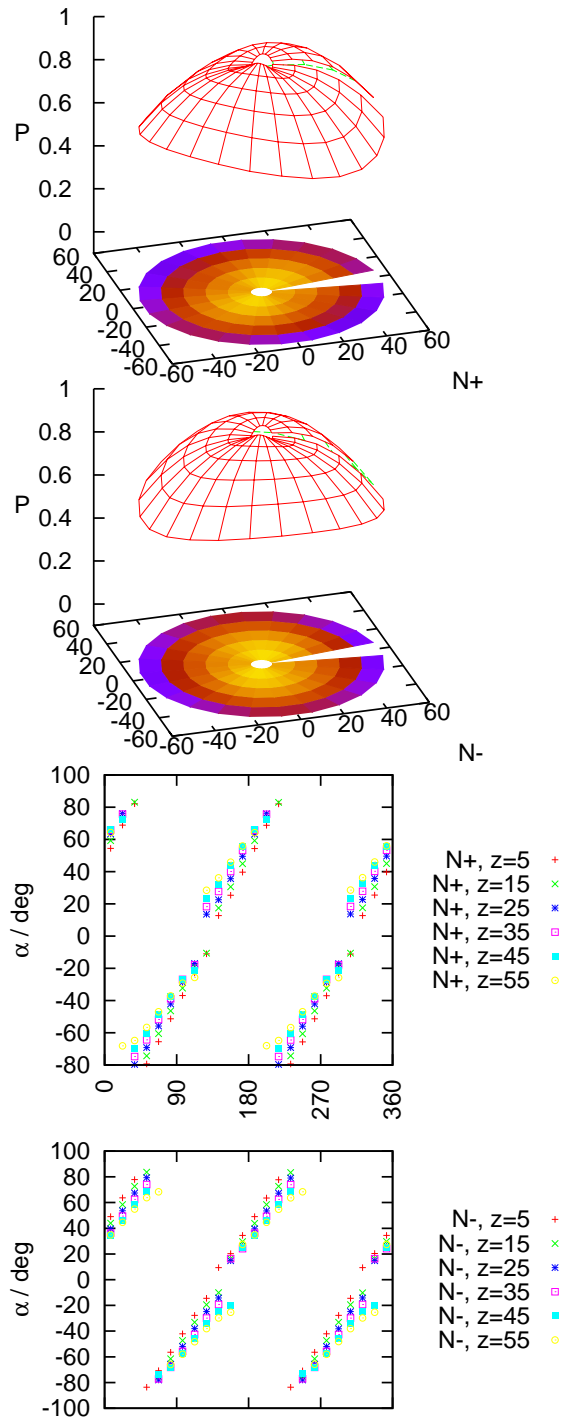}
\caption{The diattenuation (\ref{eq.P}) and angle $\alpha$ (\ref{eq.alph}) in degrees as a function of telescope pointing
at fixed rotator angle $r=0$.
}
\label{fig.pol_0}
\end{figure}

\begin{figure}[hbt]
\includegraphics[scale=0.65]{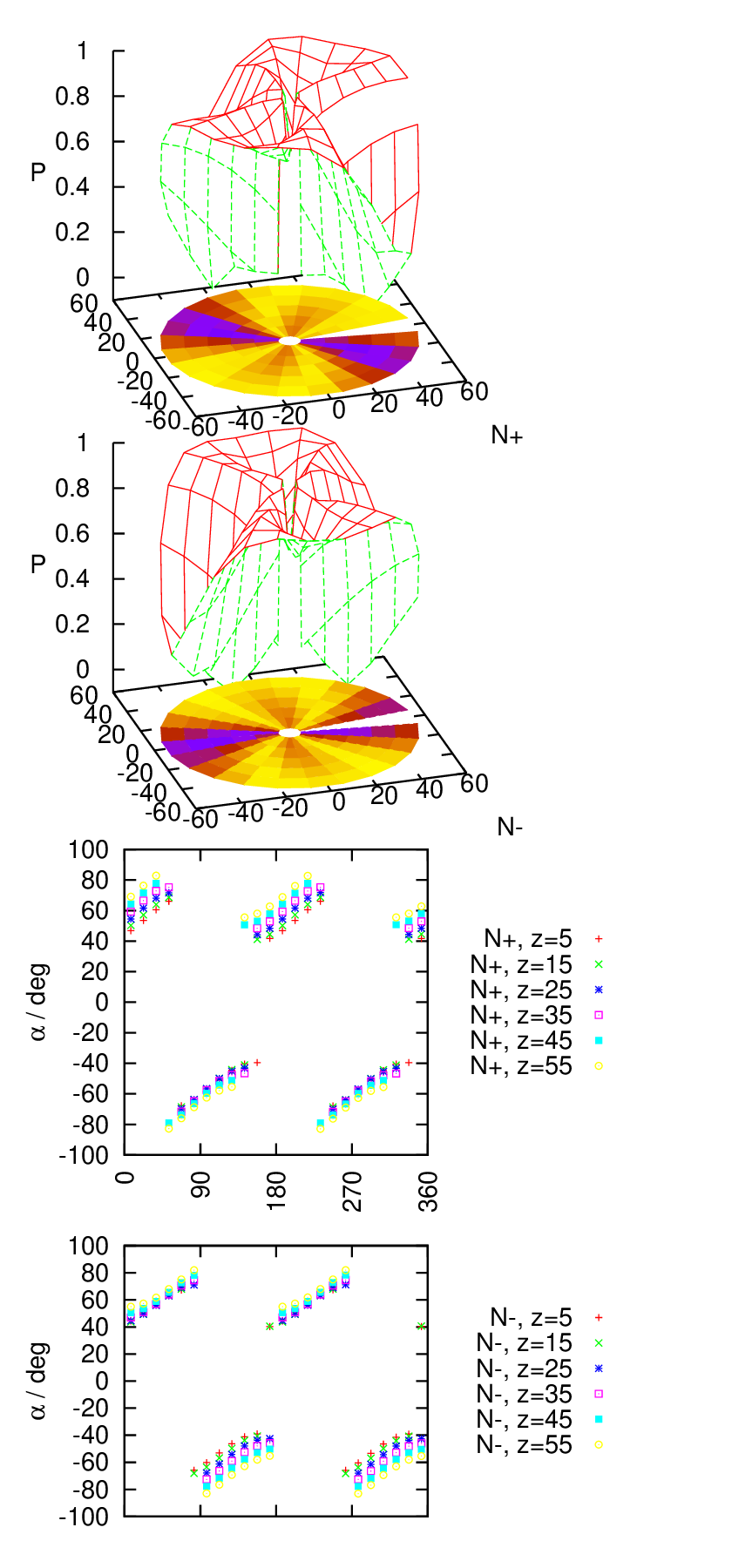}
\caption{The diattenuation (\ref{eq.P}) and angle $\alpha$ (\ref{eq.alph}) as a function of telescope pointing
at fixed rotator angle $r=30^\circ$. At azimuths $A$ for which
$P=0$, keeping $\alpha$ along the longer axis of the ellipse
creates artificial jumps of 90$^\circ$ in $\alpha$.
}
\label{fig.pol_30}
\end{figure}

\begin{figure}[hbt]
\includegraphics[scale=0.65]{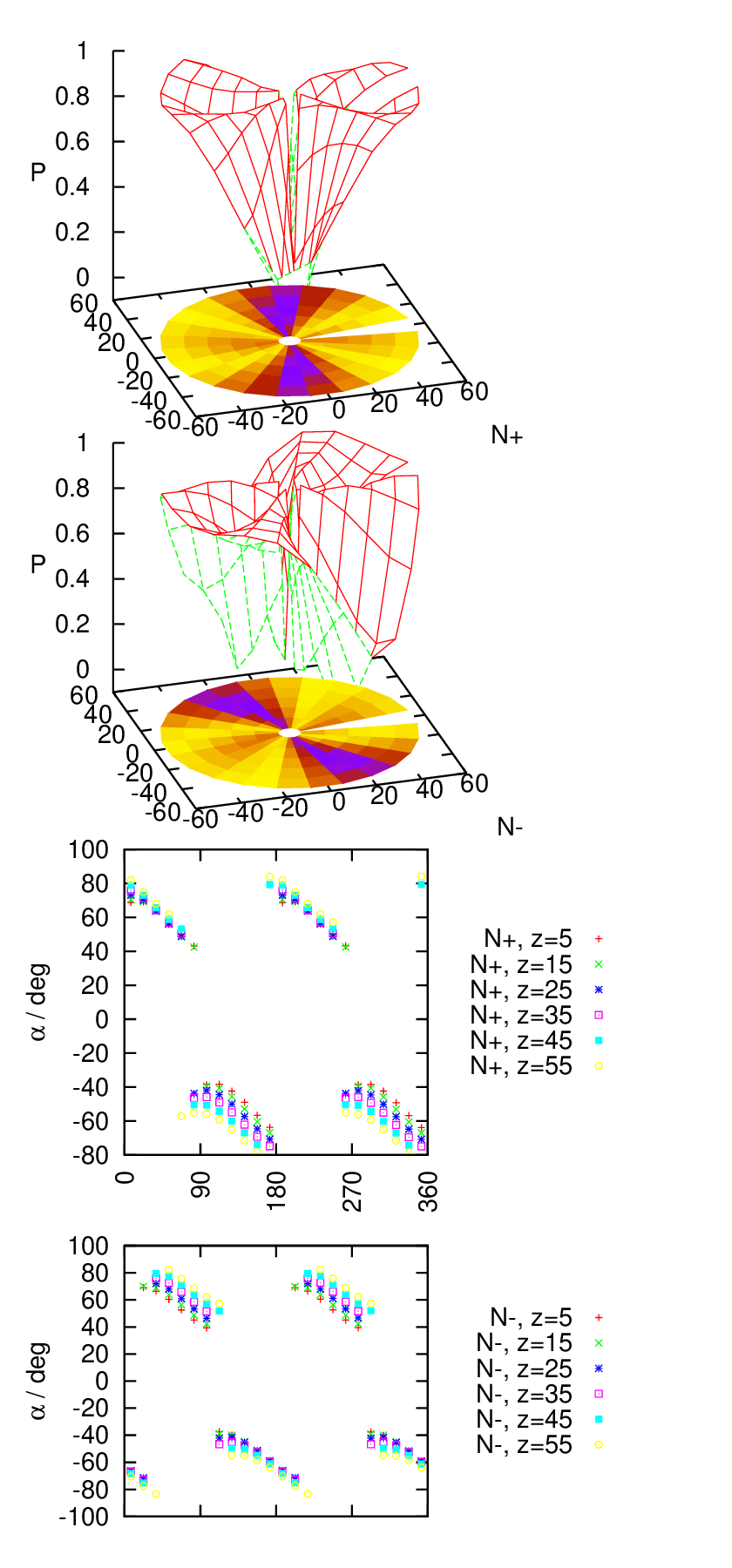}
\caption{The diattenuation (\ref{eq.P}) and angle $\alpha$ (\ref{eq.alph}) as a function of telescope pointing
at fixed rotator angle $r=60^\circ$.
}
\label{fig.pol_60}
\end{figure}

Figures \ref{fig.pol_0} to \ref{fig.pol_60} show the
influence of the pointing direction $(A,z)$ in the topocentric horizontal system and 
of the turning
angle $r$ of the D1--D3 triple on the axis ratio $P$ and on the tilt
angle $\alpha /^\circ$ of initially circularly polarized star light after its
turn into the $+V$ direction
inside the laboratory. Two plots of each figure deal with
the N+ configuration, the two other plots with the N- configuration.

In the plots of $P(A,z)$, the pointing direction is encoded in the
circular coordinates of the base: the azimuth $A=0$ is
indicated with a blank triangular sector. Pointing to the zenith, $z=0$--$5^\circ$,
happens in the middle of the small blank center. Five circular segments,
each spanning a zenith interval of 10$^\circ$, lead to
the outer rim of the plot where the zenith angle $z$ reaches 55$^\circ$.
The value of $P$
is plotted as a function over
these $(A,z)$ coordinates and in addition shown on
color scales (dark blue for small $P$, light orange for large $P$)
inside the
base circle
spanned by $A$ and $z$.
The value of $\alpha$ is shown as a scatter plot as a function of $A$ for six
different zenith angles $z$.

Each of the figures starting with Fig.\ \ref{fig.pol_0} remains unchanged if
one switches between the $+U$ and the $-U$ position of
the mirrors in the main delay line tunnel (mirrors of the main delay line
plus the associated M12 and M16 in the old nomenclature).

\clearpage

In overview, each of the two constellations N+ or N- of the Nasmyth 
station has two azimuth directions with preferred, low asymmetry
$P$ inside the beam, indicated by two
blue-shaded
radial sections
on the base of the 3D plots.
These directions co-rotate with the rotator angle $r$,
which demonstrates that ``in principle'' this degree of operational freedom
could be used 
to reduce polarization effects. In practice, the rotator angle $r$
will often be deployed to serve other needs,
which are not a topic of this manuscript.

In Figure \ref{fig.pol_0}, the rotator mirror D2 is aligned with the other
mirrors of the Coud\'e train, and the main observation is that the asymmetry
$P$ decreases while increasing the zenith angle---an effect of 
decreasing incidence angles on M3\@.

The second, main observation is that $P$ is large in the other two, unfavorable
pointing directions $A$,
whereas
the dependence on the zenith angle
is comparatively weak.

Third, the difference between the $N+$ and $N-$ configurations
for any prescribed pointing direction is smaller than one might
have hoped, although there is some influence on $P$.

\subsection{Mueller Matrix}
Another view on the same data is taken by transforming the Jones matrices
into Mueller matrices \cite{BrousseauJOSAA10,SimonJMOP34}.

Results referring to the $Q$ and $U$ components of the Stokes vector
depend on the choice of the coordinate axes of the $s$- and  $p$-polarization.
Because the mirror train is axially symmetric with respect to reflections
off M1  and M2, the $p$-direction above M1 has been defined relative to
the plane of incidence on M3, the first mirror to break this symmetry.
As we are co-rotating the Nasmyth focus with the azimuth pointing
angle $A$, the
Cartesian axes of the polarization directions above M1 are
therefore defined not a in a coordinate system laid out by the
architecture of the duct and main delay line, but 
in the coordinate system spanned by M2, M3 and M4\@.
The direction of positive $p$ is parallel to the direction
from M3 to M4 of the N+ configuration.
The main intent of this comoving polarization frame in the input pupil is
to avoid steps in the definition for azimuth angle transits through
some fixed value; as a disadvantage, the representation
contains an artificial smooth change of this frame through $360^\circ$
for one full rotation around the vertical axis.

The components in the interferometric laboratory
refer to the natural horizontal coordinate system.
(The unfortunate standard names
$U$, $V$ and $W$ of the global coordinate system used in Figs.\ \ref{fig.ATfig1}--\ref{fig.coord}
have nothing to do with these assignments of polarization states.)

\begin{figure}[hbt]
\includegraphics[scale=0.65]{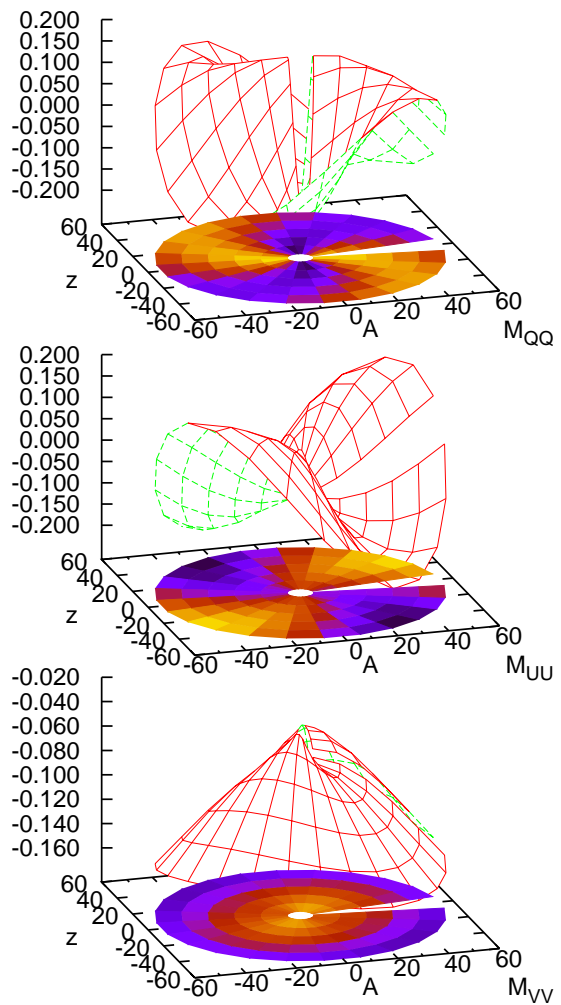}
\caption{Three diagonal
elements of the Mueller matrix
as a function of telescope pointing
at fixed rotator angle $r=0^\circ$, Nasmyth configuration N+.
}
\label{fig.mD0pW}
\end{figure}

Figure \ref{fig.mD0pW} indicates a high survival rate of star light polarization:
the three relevant diagonal elements of the Mueller 
matrix have coupling amplitudes of the order
of the intensity transmission of $\approx 0.2$.

\begin{figure}[hbt]
\includegraphics[scale=0.65]{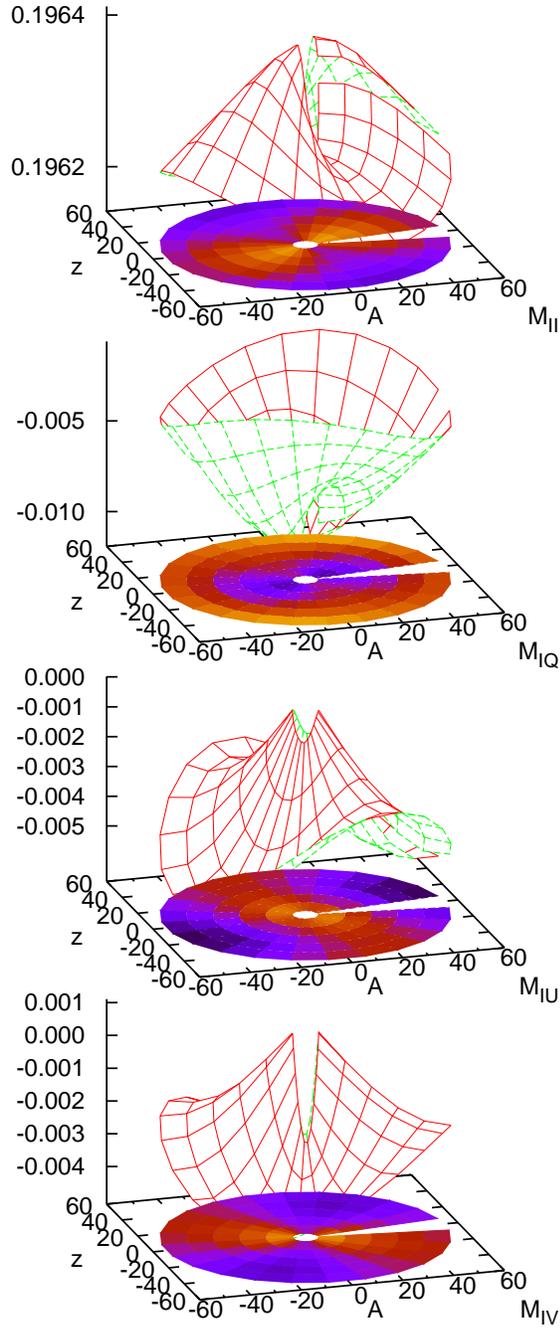}
\caption{The
four elements of the top row of the Mueller matrix
as a function of telescope pointing
at fixed rotator angle $r=0^\circ$, Nasmyth configuration N+.
}
\label{fig.m0pW}
\end{figure}

\begin{figure}[hbt]
\includegraphics[scale=0.65]{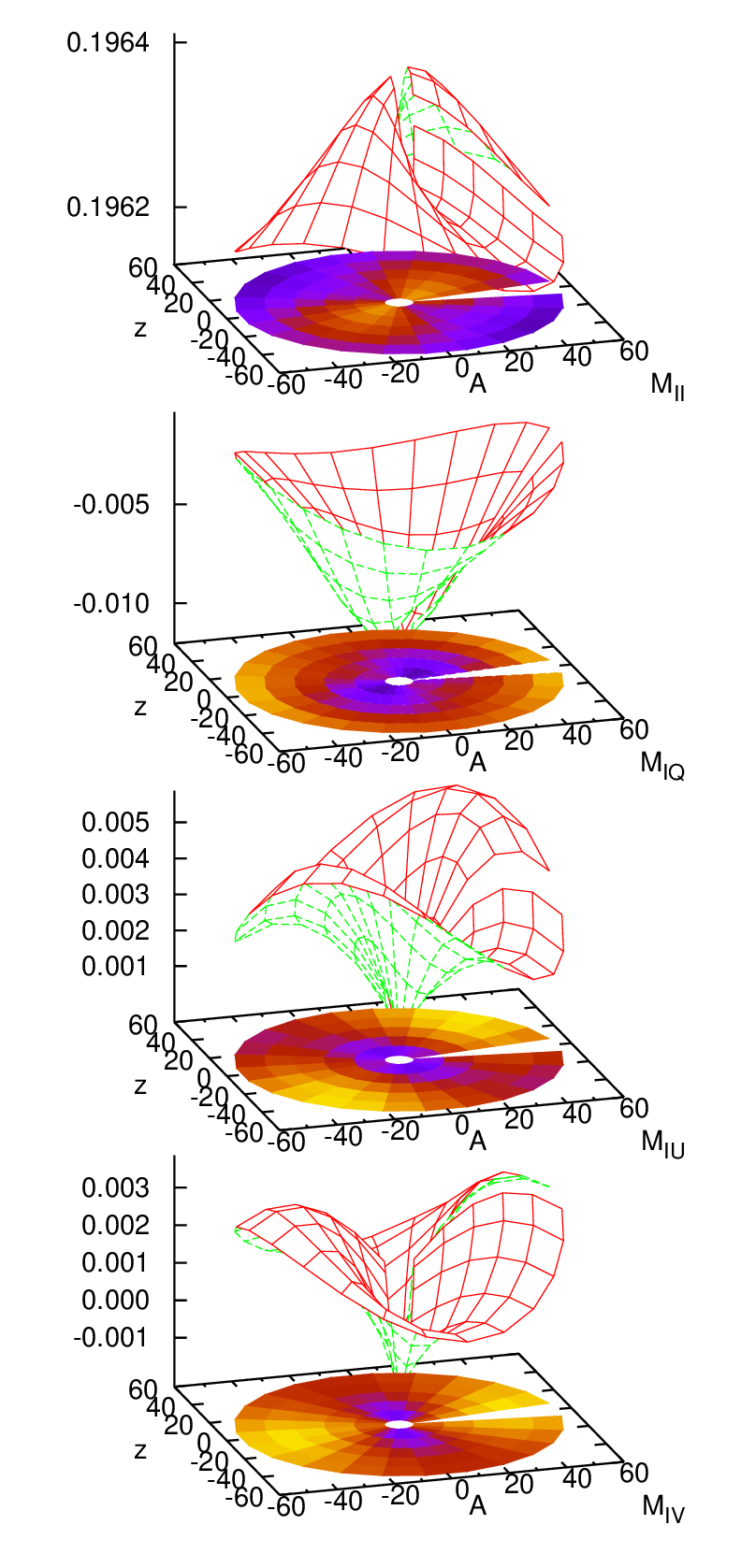}
\caption{The elements of the top row of the Mueller matrix
at fixed rotator angle $r=0^\circ$, Nasmyth configuration N-.
}
\label{fig.m0mW}
\end{figure}

\begin{figure}[hbt]
\includegraphics[scale=0.65]{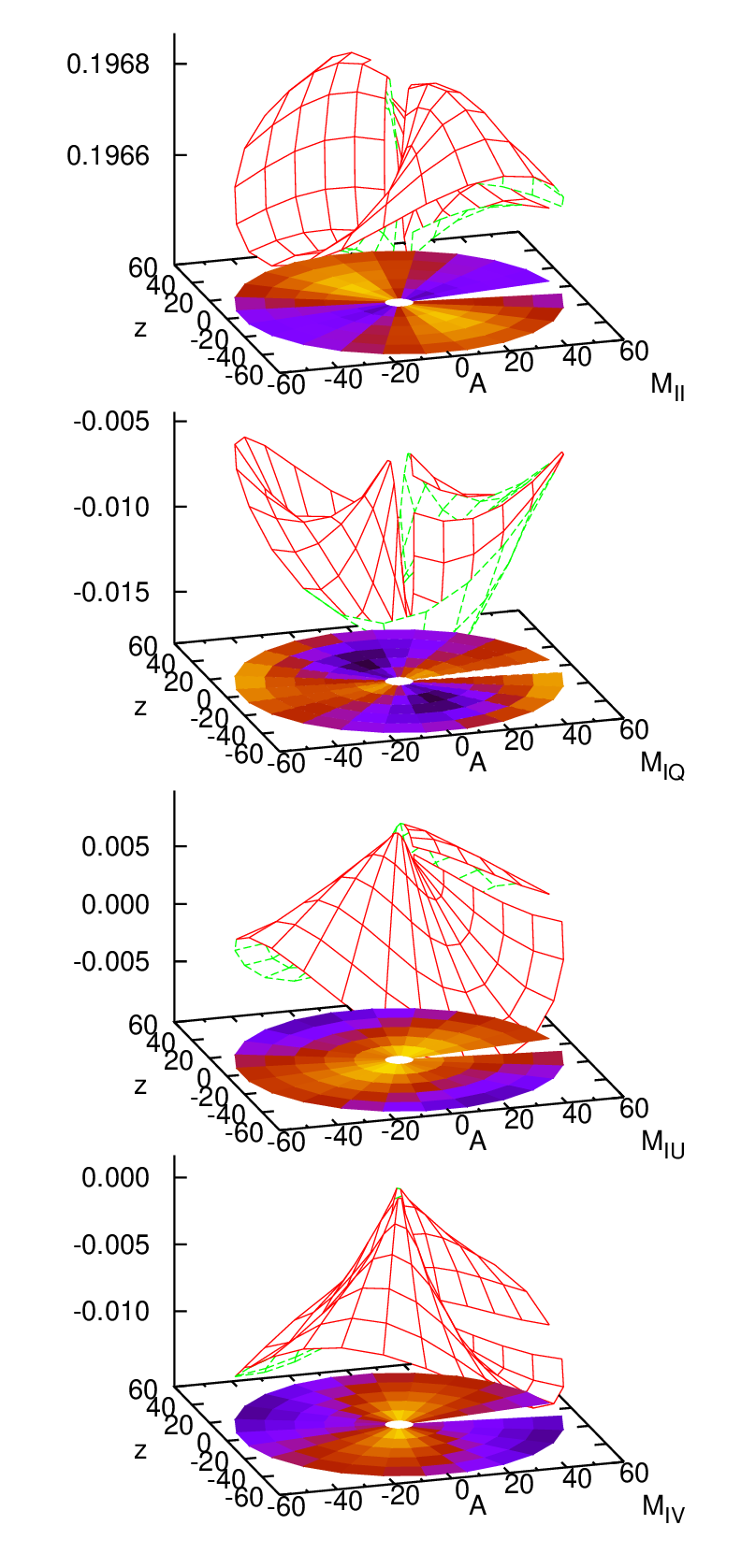}
\caption{The
elements of the top row of the Mueller matrix
at fixed rotator angle $r=30^\circ$, Nasmyth configuration N+.
}
\label{fig.m30pW}
\end{figure}

\begin{figure}[hbt]
\includegraphics[scale=0.65]{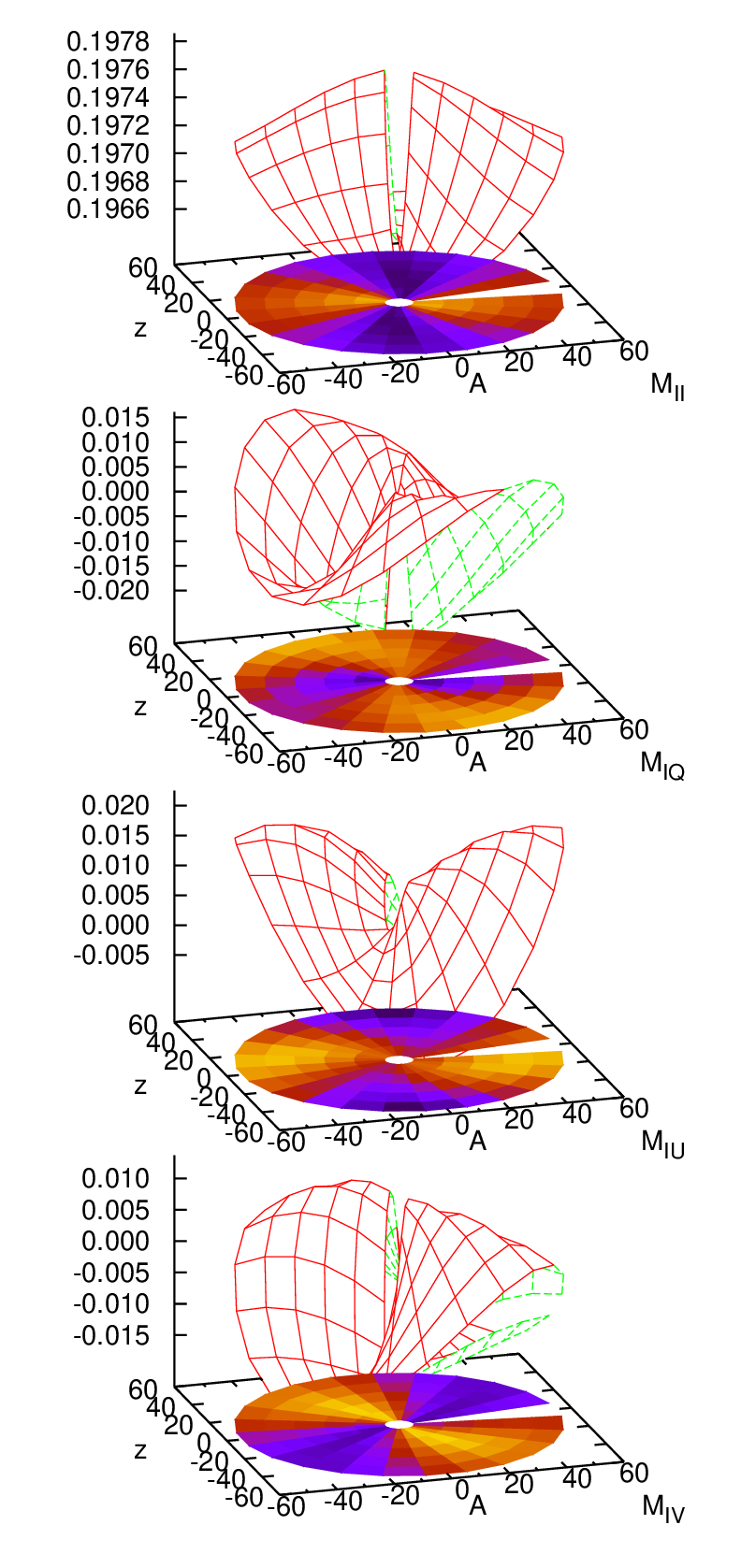}
\caption{The 
elements of the top row of the Mueller matrix
at fixed rotator angle $r=60^\circ$, Nasmyth configuration N-.
}
\label{fig.m60mW}
\end{figure}

Figures \ref{fig.m0pW}--\ref{fig.m60mW}
illustrate with the top row elements the conversion of the
$I$, $Q$, $U$ and $V$ components of the Stokes Vector into an $I$
component on exit of the mirror train.
These quantities answer the question how far a preferred state of 
polarization of the beam above M1 induces changes
in intensity observed in the interferometric laboratory.
Their top graph is $M_{II}$, the intensity-to-intensity
conversion by the mirror train, which was forced
to a value near $0.2$ by the scaling explained in Section \ref{sec.n}.
The values of the off-diagonal $M_{IQ}$, $M_{IU}$ and $M_{IV}$ in the other graphs 
vary typically over a range $\pm 0.01$, which indicates that
fully polarized star light may lead to \emph{relative} intensity variations
of up to 10\%.

As already apparent in Figs.\ \ref{fig.pol_0}--\ref{fig.pol_60},
positioning of the rotator in its ``neutral'' direction $r=0$ leads
to the least sensitivity of the
matrix elements on the pointing direction:
The spread of the $M$-values in Figs.\ \ref{fig.m0pW}--\ref{fig.m0mW}
is smaller than the spread in Figs.\ \ref{fig.m30pW}--\ref{fig.m60mW}.

\clearpage

\begin{figure}[hbt]
\includegraphics[scale=0.65]{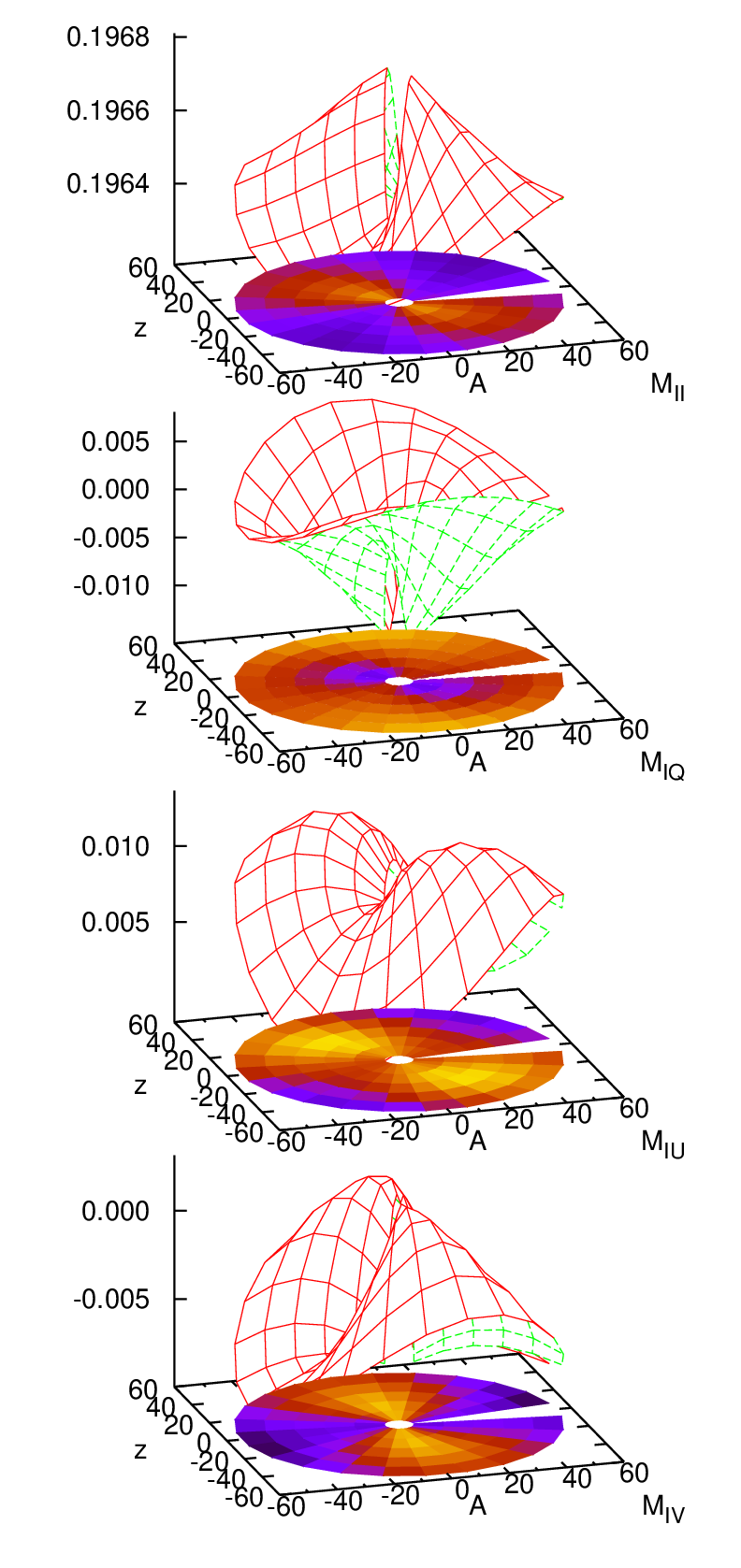}
\caption{The 
elements of the top row of the Mueller matrix
at fixed rotator angle $r=30^\circ$, Nasmyth configuration N-.
}
\end{figure}
\clearpage

\begin{figure}[hbt]
\includegraphics[scale=0.65]{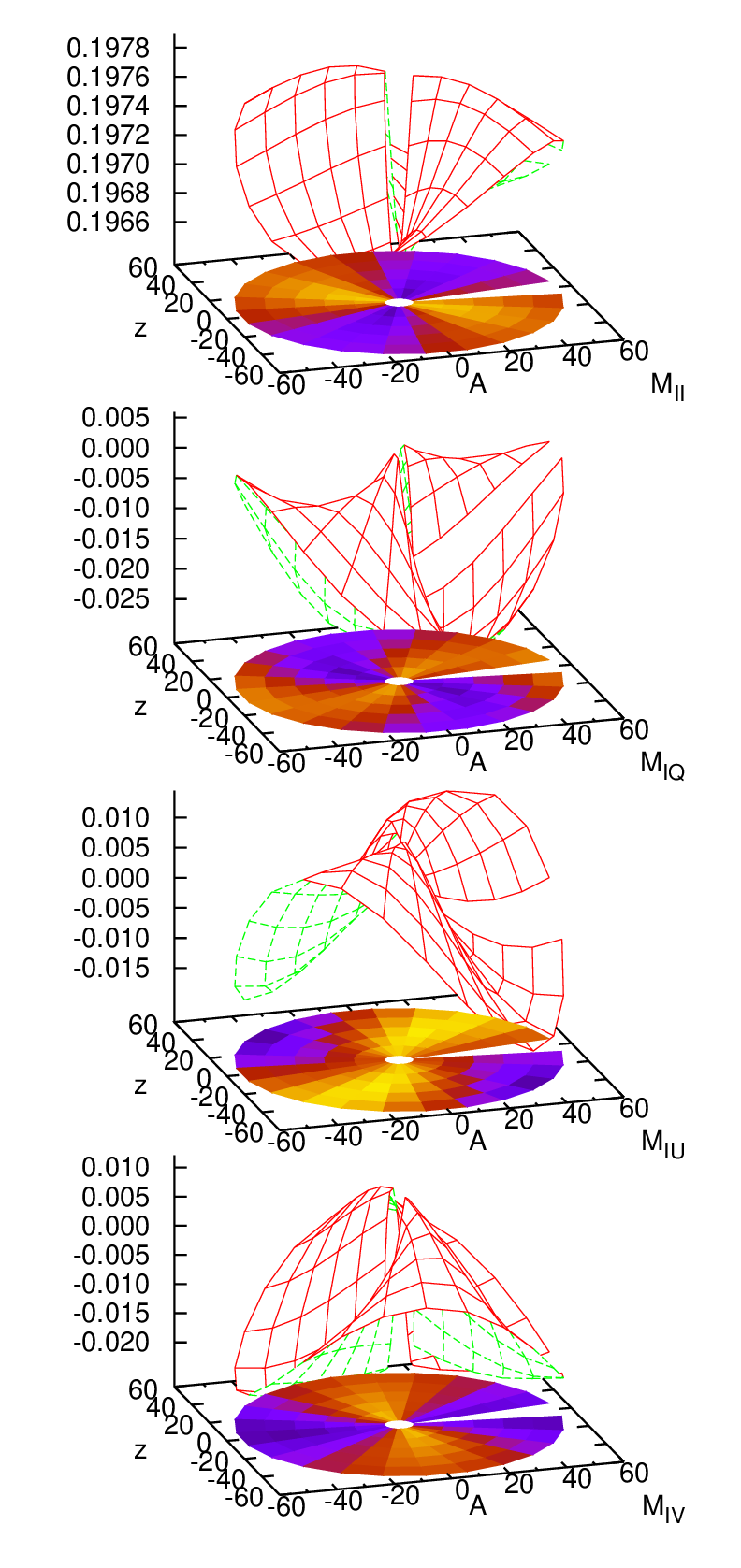}
\caption{The
elements of the top row of the Mueller matrix
at fixed rotator angle $r=60^\circ$, Nasmyth configuration N+.
}
\end{figure}

\clearpage

\section{Summary} 
Apparently this is the first publication to assess
the VLTI beam polarimetry induced by reflections off the
31 mirrors of a standard optical configuration.

The results remain semi-quantitative because they are based on a blind
estimate of the dielectric function of the mirror surfaces,
so only the dependence on
the variable parts of the mirror train geometry has been emphasized.
To first order, interferometry with a symmetric setup,
sensitive to \emph{differential} properties
of two telescope beams,
is not affected, and the calibration procedure selecting
calibrator sources nearby the science target will wipe out
most of the pointing dependencies that were outlined here.

For circularly polarized star light, the diattenuation (visibility
contrast tested with a rotating analyzer) of the beam is predicted to
reach values up to 1.0
for
unlucky pointing azimuths, which are themselves a function
of the star rotator position.

\bibliographystyle{apsrmp}
\bibliography{all,eso}

\end{document}